\documentclass[12pt]{article}
\usepackage[left=2cm, right=2cm, bottom=2cm ,top = 2cm, footskip=1cm]{geometry}
\usepackage[font=small,labelfont=bf]{caption}

\usepackage{setspace}
\usepackage{lineno}

\let\counterwithin\relax
\usepackage{chngcntr}

\usepackage{hyperref}
\usepackage{subcaption}
\usepackage{titling} 
\newcommand{\subtitle}[1]{%
  \posttitle{%
    \par\end{center}
    \begin{center}\Large#1\end{center}
    \vskip0.1em}%
}

\usepackage[authoryear,sort&compress]{natbib}
\usepackage{multibib}
\newcites{supp}{Supplementary References}
\setcitestyle{open={(},close={)}}

\usepackage{graphicx}
\usepackage{amsmath}

\usepackage[section]{placeins}
\title{Dual mechanism of Anti-Seizure Medications in controlling seizure activity}

\author{Guillermo M. Besn\'e$^{1}$, Emmanuel Molefi$^{1}$, Sarah J. Gascoigne$^{1}$, 
\\Nathan Evans$^{1}$, Billy Smith$^{1}$, Chris Thornton$^{1}$, \\ 
Fahmida A. Chowdhury$^{3}$, Beate Diehl$^{3}$, John S. Duncan$^{3}$,\\ 
Andrew W. McEvoy$^{3}$, Anna Miserocchi$^{3}$, Jane de Tisi$^{3}$, Matthew Walker$^{3}$,\\ , Peter N. Taylor$^{1,2,3}$, Yujiang Wang$^{1,2,3*}$}

 \usepackage{float}
\usepackage{aliascnt}
\newaliascnt{eqfloat}{equation}
\newfloat{eqfloat}{h}{eqflts}
\floatname{eqfloat}{Equation}

\newcommand*{\MyIndent}{\hspace*{0.5cm}}%
\newcommand*{\ORGeqfloat}{}
\let\ORGeqfloat\eqfloat
\def\eqfloat{%
  \let\ORIGINALcaption\caption
  \def\caption{%
    \addtocounter{equation}{-1}%
    \ORIGINALcaption
  }%
  \ORGeqfloat
}

\begin{document}

\maketitle

\begin{enumerate}
\item{CNNP Lab (www.cnnp-lab.com), School of Computing, Newcastle University, Newcastle upon Tyne, United Kingdom}
\item{Faculty of Medical Sciences, Newcastle University, Newcastle upon Tyne, United Kingdom}
\item{UCL Queen Square Institute of Neurology, Queen Square, London, United Kingdom}
\end{enumerate}

\begin{center}
* Yujiang.Wang@newcastle.ac.uk    
\end{center}

\newpage

\section*{Abstract}
\textbf{Background:} Anti-seizure medications (ASMs) can reduce seizure duration, but their precise modes of action are unclear. Specifically, it is unknown whether ASMs shorten seizures by simply compressing existing seizure activity into a shorter time frame or by selectively suppressing certain seizure activity patterns.

\textbf{Methods:} We analysed intracranial EEG (iEEG) recordings of 457 seizures from 28 people with epilepsy undergoing ASM tapering. Beyond measuring seizure occurrence and duration, we categorized distinct seizure activity patterns (states) based on spatial and frequency power characteristics and related these to different ASM levels.

\textbf{Results:} We found that reducing ASM levels led to increased seizure frequency ($r = 0.87, \ p < 0.001$) and longer seizure duration ($\beta = -0.033, \ p < 0.001$), consistent with prior research. Further analysis revealed two distinct mechanisms in which seizures became prolonged:

Emergence of new seizure patterns – In approx. 40\% of patients, ASM tapering unmasked additional seizure activity states, and seizures containing these ``taper-emergent states'' were substantially longer ($r = 0.49, \ p < 0.001$).

Prolongation of existing seizure patterns – Even in seizures without taper-emergent states, lower ASM levels still resulted in approx. 12-224\% longer durations depending on the ASM dosage and tapering ($\beta = -0.049, \ p < 0.001$).

\textbf{Conclusion:} ASMs influence seizures through two mechanisms: they (i) suppress specific seizure activity patterns (states) in an all-or-nothing fashion and (ii) curtail the duration of other seizure patterns. These findings highlight the complex role of ASMs in seizure modulation and could inform personalized dosing strategies for epilepsy management. These findings may also have implications in understanding the effects of ASMs on cognition and mood.

\section*{Keywords}

neurophysiology, dynamic states, seizure states, seizure propagation, bifurcation


\newpage

\section{Introduction}

Anti-seizure medications (ASMs) are a cornerstone of epilepsy treatment, helping many patients achieve seizure control~\citep{Kanner2022}. However, determining the right dosage remains a challenge, as treatment must balance seizure suppression with potential side effects. Currently, ASM dosing follows a trial-and-error approach, tailored to each patient’s unique response. While seizure suppression is generally dose-dependent, ASMs can reach a ceiling effect, beyond which increasing the dose provides no additional benefit~\citep{Panayiotopoulos2004}. In fact, approximately 30\% of epilepsy patients continue to experience seizures despite trying multiple ASMs at increasing doses~\citep{Ramgopal2013}. Although the molecular and cellular mechanisms of ASMs have been well studied~\citep{Mula2021,Dini2023,Clemens2006,Sarangi2022,Kanner2022}, their dose-dependent effects on human brain activity and seizure dynamics \textit{in vivo} remain poorly understood. 

Previous studies suggest that higher ASM doses are associated with shorter seizures~\citep{Ghosn2023,Holler2018,Sarangi2022}. However, the mechanisms underlying this effect are unclear -- ASMs may simply compress seizure activity into a shorter timeframe, or actively suppress the involvement of specific brain regions and activity patterns in an all-or-nothing fashion. In this study, we tested if either of these two mechanisms dominate, or if a mixture of both mechanisms exist. 

To this end, we retrospectively analysed intracranial EEG (iEEG) data from patients undergoing pre-surgical epilepsy monitoring, where ASMs are often gradually withdrawn to trigger more severe seizures. This controlled setting provides a unique opportunity to investigate how different ASM doses and combinations influence brain activity in humans \textit{in vivo}. Building on previous work \cite{burns2014network,Schroeder2023}, we dissect seizure activity into activity patterns and therefore describe each seizure as a sequence of these patterns or states. By examining seizure states across varying ASM levels, we aim to uncover fundamental principles of drug action that may be relevant across neurological disorders, and may contribute to precision medicine approaches. 


\newpage
\section{Methods}
\label{sec:Methods}

\subsection{Subject and study information}

In this retrospective study, we included 28 subjects with medically refractory epilepsy who underwent intracranial EEG (iEEG) monitoring and medication tapering. Anonymised ictal recordings were obtained from the National Epilepsy \& Neurology Database and analysed with the approval of Newcastle University Ethics Committee (42569/2023). All available subjects from the National Epilepsy \& Neurology Database for this study were contributed by the National Hospital for Neurology and Neurosurgery (NHNN).  

For this study, we only included subjects for whom we had at least three usable seizure EEG and imaging data to allow for localising iEEG contacts to brain regions. Electrographic seizure onset and offset times were marked by expert epileptologists at the NHNN, algorithmically verified, and visually confirmed by SJG, NE, and YW. If seizures occurred in quick succession (i.e., $< 120$~s between offset and subsequent onset), only the lead seizure was included. We further excluded any seizures following sleep deprivation or cortical stimulation procedures, or occurring during periods where ASM levels exceeded the typical range (e.g., following administration of rescue medication).

\subsection{iEEG pre-processing}

All seizure EEG recordings were extracted with an accompanying 120~s pre-ictal period. Any EEG sampled at higher than 512~Hz was resampled with anti-aliasing to 512~Hz. All iEEG signals were then band-pass filtered between 0.5 and 200~Hz, and notch filtered at 50~Hz (and harmonics) using a 2~Hz window to exclude line noise. All filtering was performed using a zero-phase response 4-th order Butterworth filter. To identify pre-ictal noise, an iterative detection algorithm was used, followed by visual validation. For each subject, noisy iEEG channels were removed from all seizures. If a single seizure had many noisy channels, we opted to exclude that seizure to maintain the maximum number of channels for that subject. Finally, recordings were re-referenced to a common average.

\subsection{Localisation of electrodes to Regions of Interest (ROIs)}

iEEG electrode contacts were localised to brain regions of interest (ROIs) to provide anatomical context. Following our previously described pipeline \citep{wang2023temporal,Thorton2023}, electrodes were mapped to 120~ROIs from the Lausanne `scale60' atlas \citep{hagmannMappingStructuralCore2008} using pre-operative MRI parcellations generated via FreeSurfer \citep{hagmannMappingStructuralCore2008, fischlFreeSurfer2012}. Electrode contacts were assigned to the nearest grey matter region within 5~mm, with those beyond this distance excluded from analysis. \cite{Woodhouse2025} expatiates on the pipeline used herein.

\subsection{Identification of seizure states}\label{METH_2.4}

\begin{figure}[H]
    \centering
    \includegraphics[scale=1]{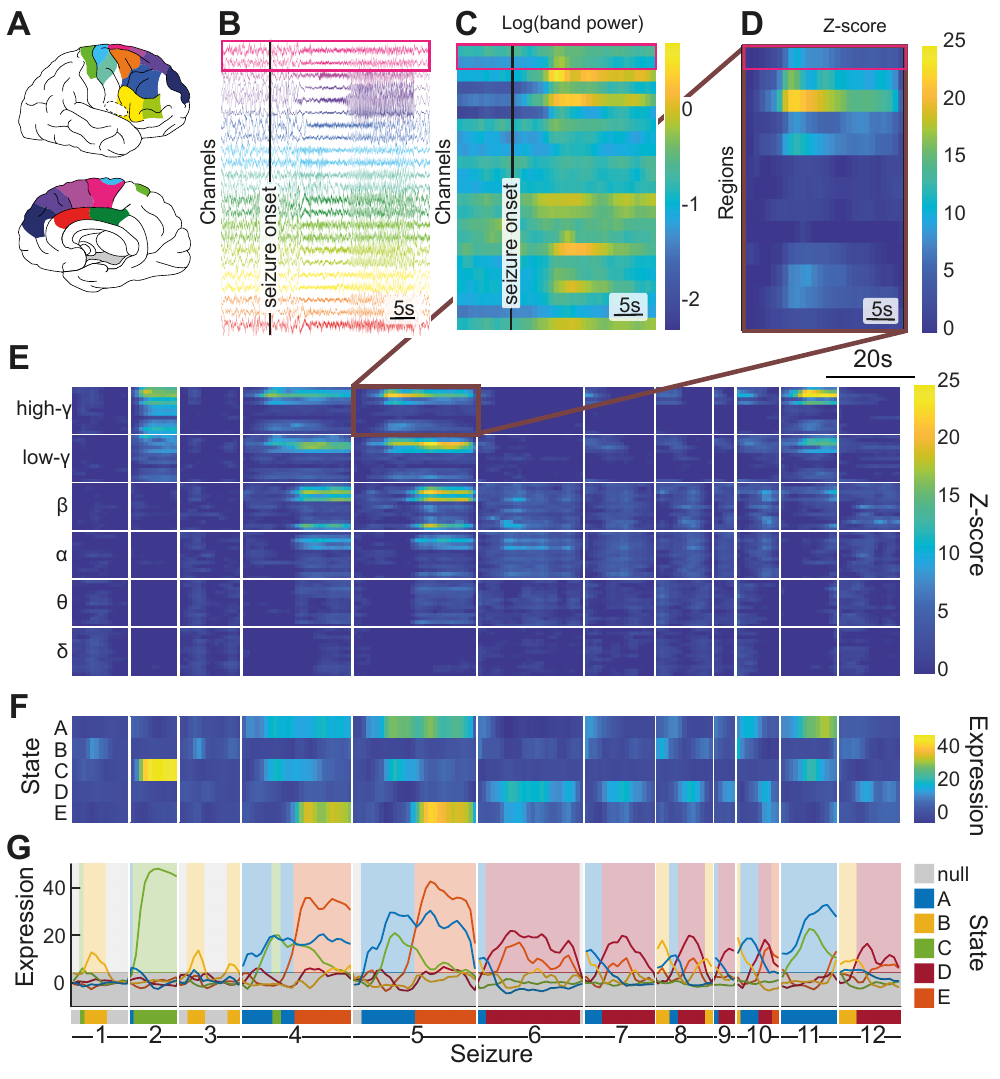}
    \caption{\textbf{Seizure state identification from EEG for one example subject.}
        \textbf{A and B} Implanted brain regions and associated EEG channels.
        \textbf{C} Log power in high-$\gamma$ band in over time windows of 5~s with 4~s overlap in all EEG channels.
        \textbf{D} Z-scored log power of ictal \textit{vs.} pre-ictal period in high-$\gamma$ band averaged within each brain region.
        \textbf{E} Data matrix $X$ consisting of z-score matrices stacked across frequency bands and concatenated across seizures.
        \textbf{F} Expression matrix $H$ following non-negative matrix factorisation of $X$.
        \textbf{G} State identification using dominant component in $H$ at each time window.
    }
    \label{Methods1}
\end{figure}

To characterise subject-specific seizure activity patterns, we grouped time periods within all seizures of an individual into `seizure states' \citep{Burns2014,Schroeder2023}. Figure~\ref{Methods1} demonstrates the workflow from iEEG (panel A and B) to state allocation (panel G). Supplementary~\ref{SUP_2} provides additional algorithmic details.

For each seizure in every subject, we computed band power using a 5~s sliding window with 4~s overlap in six distinct frequency bands ($\delta$: 1-4~Hz , $\theta$: 4-8~Hz, $\alpha$: 8-13~Hz, $\beta$: 13-30~Hz, low-$\gamma$: 30-60~Hz, and high-$\gamma$: 60-100~Hz). Figure~\ref{Methods1}A and B show an example seizure for an example subject, with implanted brain regions colour-coded. Band power data were log-transformed (Figure~\ref{Methods1}C shows the high $\gamma$ band), and then averaged across recording channels in each brain region. We z-scored band power values from seizure time windows to the pre-ictal period (Figure~\ref{Methods1}D). Thus, for each seizure, and each frequency band, we obtained a matrix with dimensions $n_{regions} \times n_{T}$, where $n_{regions}$ and $n_{T}$ denote the number of recorded regions, and of time windows in a given seizure, respectively.

For each subject, we then stacked and concatenated all such matrices to create data matrix $X$ (Figure~\ref{Methods1}E) with dimensions $(n_{regions}*n_{bands}) \times
n_{T_{all}}$, where $n_{bands}=6$ is the number of frequency bands, and $n_{T_{all}}$ the number of time-windows across all seizures in a given subject.

To obtain seizure states, we used non-negative matrix factorisation (NMF) to decompose $X$ into two non-negative matrices ($X \approx W \times H$), such that the information in $X$ is summarised by a few components. Here, how strongly each component is expressed at any given time window is captured in $H$, which has dimensions $n_c \times n_{T_{all}}$ with $n_c$ being the number of components. Figure~\ref{Methods1}F shows $H$ with $n_c=5$ in our example subject. $W$ has dimensions $(n_{regions}*n_{bands}) \times n_c$ and captures the patterns of seizure activity in terms of frequency band and brain regions. We can interpret these as spatio-frequency patterns of seizure activity. Together, $W$ tells us what patterns of seizure activity are present, and $H$ tells us how strongly these are expressed at any given time window.

We used $H$ to assign each time window to a component (denoted herein as ``state'') by selecting the most strongly expressed component. This simplistic approach effectively turns the soft-clustering provided by NMF into a hard clustering, and we added some thoughts for future work in our~\nameref{sec:Discussion} on this choice. In some time windows, none of the components are particularly strongly expressed, indicating that there is generally a low level of seizure activity. We therefore thresholded the expression level (Figure~\ref{Methods1}G, grey area) and assigned time windows without strong expression to a ``null'' state. This step means that we are primarily capturing seizure propagation patterns, and ignoring subtle seizure onset patterns (see~\nameref{sec:Discussion}). In Figure~\ref{Methods1}G, we show the final identified states in all seizures in our example subject.



\subsection{ASM tapering and plasma concentration estimation}

The modelling of ASM plasma concentration described in our previous study \citep{Besne2024} is replicated here with some modifications. Each subject's ASM dosage schedule was obtained from clinical records then converted into a continuous estimation of plasma concentration using known pharmaco-kinetics \citep{Ghosn2023,DB,Patsalos2022,Iapadre2018}. It is important to note that the precise timing of each intake is not available, and provided is only the schedule of intake. To retain the natural daily fluctuations of the medication and reduce the effects of this lack of accuracy, we applied a 6~h moving average with a 5~h overlap. 

\begin{figure}[H]
    \centering
    \includegraphics[scale=1]{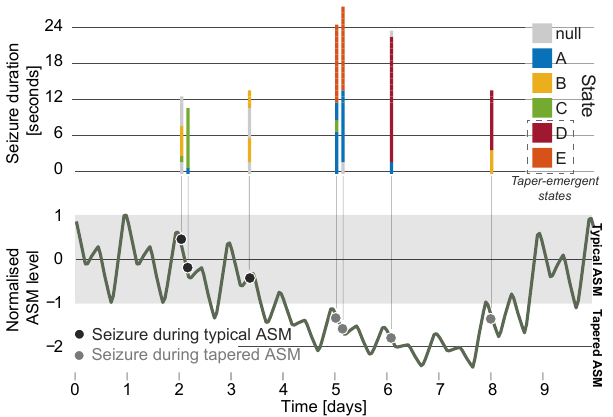}
    \caption{\textbf{ASM level and seizure states for an example subject.}
        Top: Seizure states for each seizure are plotted as a colour-coded bar in time. Y-axis indicates duration of each seizure. Bottom: Normalised ASM levels over recording time with the `typical' ASM condition (between $\pm$1) highlighted in grey. Seizures are marked as dots and aligned with top panel. Seizure states D and E are ``taper-emergent states'' as they only appear in seizures that occurred during the tapered ASM condition.
    }
    \label{Methods2}
\end{figure}

The variability in pharmacokinetics and dosage, results in a different range of blood plasma concentration (BPC) across subjects and ASMs. However, the range of fluctuations before tapering should have similar therapeutic effect across subjects \citep{Besne2024,Ghosn2023,Flanagan2008}. To allow for comparison, we normalised plasma concentration values into a single ``ASM level" that would fluctuate between $\pm$1 during non-tapered (typical ASM) periods for each subject. Initially, each ASM plasma concentration was normalised individually using 

\begin{equation}
    \label{BPCnorm}
    BPC_{norm} = {\displaystyle \frac{BPC - mean(BPC_{steady})}{range(BPC_{typical})/2}},
\end{equation}

where $BPC_{typical}$ is all blood plasma concentration from the non-tapered typical medication regime at the start of the recording. A combined plasma concentration was then calculated as a mean of all medications a subject was taking and re-normalised using the same equation, i.e., Equation~\ref{BPCnorm}. Figure~\ref{Methods2}~(bottom panel) presents the resultant normalised ASM level estimation in an example subject. This normalised ASM level will be used throughout the remaining paper. 

We then algorithmically defined `typical' and `tapered' ASM periods in each subject according to thresholds: Typical periods represent the expected ASM levels according to the subject's treatment regimen prior to tapering. Periods with ASM levels in the range of $\pm$1 were thus labelled as `typical'. We labelled periods with ASM levels below $-1$ as `tapered'. As previously mentioned, periods where ASM levels exceeded a value of 1 (rescue medication) were excluded. Figure~\ref{Methods2}~(bottom panel) provides a visual summary of this definition in an example subject.


Finally, within each subject, we identified seizure states that only occurred in seizures during the tapered condition (if any existed), and term them ``taper-emergent states''. These represent seizure activity patterns in certain frequency bands and brain regions that are only seen in seizures during the tapered condition. Figure~\ref{Methods2}~(top panel) shows this in our example subject. Note that in contrast to these ``taper-emergent states'', we refer to all other states in a given subject as ``dose-independent states'' given these can occur independent of ASM condition.

\subsection{Statistical analyses}

For our analysis, we performed a log10 transformation of seizure duration, determined from clinically labelled seizure onset and offset times. This choice was guided by previously described methods \citep{Ghosn2023,Schroeder2023} due to the distribution of seizure duration generally being heavy-tailed. The log transformation effectively means that we are comparing relative seizure duration. For example, if seizure log duration decreased by --0.2 between two conditions, this can be translated as a 63\% ($10^{-0.2}= 0.63$) reduction to duration. We report these percentage reductions throughout the~\nameref{sec:Results}.

Using the whole dataset, we compared seizure frequency and duration across different ASM levels. Seizure frequencies during typical and tapered periods were compared using a paired-sample Wilcoxon rank-sum test. We report the $r$-value and associated $p$-value.

We then related the continuous measures of ASM level and duration of seizures using a linear mixed-effects model (LMM) with the following formula: 

\begin{equation}
    \begin{split}
    log{\text -} duration \sim &  \beta_0 \\
    & + \beta_1ASM{\text -}level \\
    & + \beta_2sin(24hT) + \beta_3cos(24hT) \\
    & + \beta_4sin(12hT) + \beta_5cos(12hT) \\
    &+ (1|subject)
    \end{split}
\label{MELM}
\end{equation}

where $subject$ is included as a random intercept to control for systematic differences in seizure duration between subjects. Because seizure durations have been reported to have circadian and ultradian rhythms \citep{Xu2020,Jin2020}, we have included circular variables (sine and cosine of $12hT$ and $24hT$) to ensure that the effects modelled are not simply capturing 12 or 24~h fluctuations in seizure durations. We report the slope coefficient for ASM-level ($\beta_1$) and the associated $p$-value as a measure of the relationship between ASM-level and seizure duration, after factoring out subject-specific and chronobiological effects. 

To test for group-level effect of, e.g., tapered \textit{vs.} typical ASM conditions on seizure duration, we compared LMM-corrected durations of typical and tapered seizures using Wilcoxon rank-sum tests. The LMM-corrected durations are simply the residuals of Equation~\ref{MELM} with the ASM-level effect added back. These corrected durations represent the duration of seizures after removing subject-specific offsets and chronobiological effects. 

Throughout this work, we provide $p$-values as a reference. However, we do not use these $p$-values for down-stream analyses, and avoid the interpretation of statistical significance.

\subsection{Data and Code Availability}
Anonymised EEG band power data and ASM intake schedule, along with analysis code, will be available on GitHub: \url{https://github.com/cnnp-lab/2025_ASM-SzState_GMB}. 


\section{Results}
\label{sec:Results}

\subsection{Subject and data characteristics}

In this retrospective study, we assessed 457 seizures in 28 adult subjects who underwent iEEG monitoring and ASM tapering. Our cohort contained a mixture of temporal lobe and extra temporal epilepsies. 
Full metadata and demographics are provided in Supplementary Table~\ref{table:Demograph}. 

\subsection{ASM tapering triggers more frequent and longer seizures}\label{results1}

\begin{figure}[H]
    \centering
    \includegraphics[scale=1]{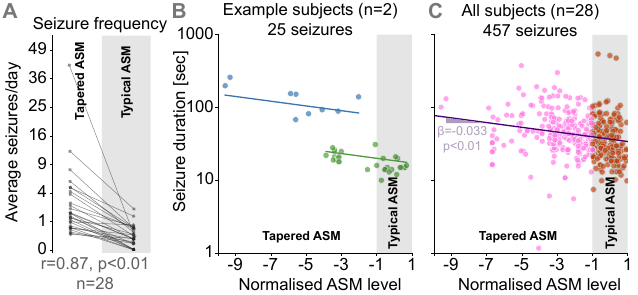}
    \caption{\textbf{Seizure frequency and duration change with ASM tapering.}
        \textbf{A} Seizure frequency measured as number of seizure per day averaged across the entire condition (`Typical ASM' or `Tapered ASM') is shown for all subjects (n=28). Wilcoxon's signed-rank test shows a substantial increase ($r=0.87$, $p < 0.001$) in seizure frequency in the tapered condition.
        \textbf{B} In two example subjects, we plotted seizure duration against normalised ASM levels in all their seizures. Lines of best fit are plotted for each subject for reference. 
        \textbf{C} For all subjects (n=28) and their 457 seizures, we plotted seizure duration after removing systematic differences between subjects (using a random offset linear mixed-effects model -- LMM) against normalised ASM levels. Seizures occurring in the typical and tapered conditions are shown in orange and pink, respectively. Best fit of fixed effects from a LMM is shown as a solid line, with a slope $\beta = -0.033$, $p < 0.001$.
    }
    \label{fig:result_asm_freq}
\end{figure}

First, to confirm findings from previous literature and validate our data, we investigated whether ASM tapering was associated with a change in seizure frequency, and duration. We compared seizure frequency, measured as seizures per day, between typical (normal ASM dosage) and tapered conditions (Figure~\ref{fig:result_asm_freq}A). Seizure frequency was substantially higher in the tapered condition than in the typical condition (Wilcoxon signed-rank test $r=0.87$, $p < 0.001$).

Finally, we asked whether seizure log-duration was correlated with ASM withdrawal. Figure~\ref{fig:result_asm_freq}B displays the log-duration against normalised ASM plasma concentrations for two example subjects. It is clear from Figure~\ref{fig:result_asm_freq}B that different subjects have systematically different seizure durations \citep[as reported in][]{Ghosn2023}, which requires a subject-level normalisation to enable group-level comparisons. To this end, we used a linear mixed-effects model with subject-level random intercepts. This allowed us to put all subjects on the same scale in terms of duration as shown in Figure~\ref{fig:result_asm_freq}C. Our analysis using a linear mixed-effects model showed a strong link between ASM withdrawal and longer seizure durations ($\beta = -0.033$, $p < 0.001$). The value $\beta = -0.033$ represents the change in log10(seizure duration) for each unit change in normalised ASM levels. This translates to an approximate 8\% increase in seizure duration for every unit reduction in normalised ASM levels. Under typical conditions (with a stable ASM dosage), ASM levels naturally fluctuate between +1 and --1 units. This means that within this normal range, seizure duration may vary by approximately 16\%. However, during ASM tapering, each unit decrease in ASM levels is expected to further increase seizure duration by about 8\%. Most subjects in our study experienced a 2- to 10-fold reduction in normalised ASM levels. Thus, we estimate that seizure duration could additionally increase by 16\% to 80\% in the tapered condition. Full results of this model are presented in Supplementary~\ref{SUP_MELM}.

Having confirmed that in our dataset, seizure duration is indeed increased with ASM tapering, we then sought to investigate how exactly the seizure EEG patterns change with ASM tapering.

\subsection{Some seizures during ASM tapering involve additional regions and activity patterns}

To understand the exact EEG patterns, we divided all the seizure EEG patterns for a given subject into a distinct number of seizure states (see~\nameref{sec:Methods}). This usually decomposed each seizure into a sequence of a few states (Figure~\ref{Methods1}).

We first explored if some states (i.e., seizure EEG patterns) only appear in the tapered condition within each subject. To achieve this, we selected subjects who had at least three seizures overall, with at least one in each of the tapered and typical ASM conditions, resulting in 304 seizures across 17 subjects. This selection step ensured that, per subject, we saw at least one seizure arising from the typical and tapered condition, respectively. We could then proceed to detect states that only appear in the tapered condition, herein termed ``taper-emergent states''. Indeed, we found 7~subjects with 30~seizures that contained taper-emergent states. On average per subject, there were $1.43 \pm 0.73$ taper-emergent states, and $4.29 \pm 2.12$ seizures that contained them. In all but one subject, these taper-emergent states recruited additional brain regions into seizure activity (Suppl.~\ref{SUP_TpaSS}).

\subsection{Seizures containing taper-emergent states are substantially longer\label{TSS}}

\begin{figure}[H]
    \centering
    \includegraphics[scale=1]{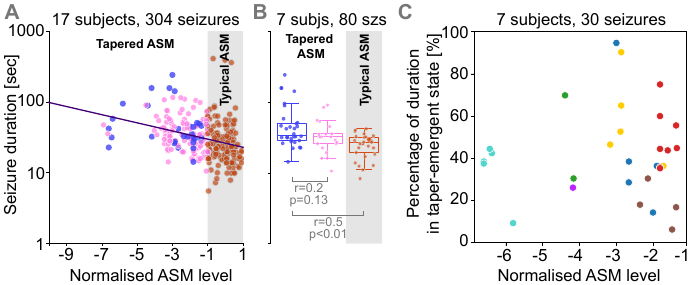}
    \caption{\textbf{Seizures containing taper-emergent states are longer, and the taper-emergent states themselves contribute substantially to the overall seizure duration.}
    \textbf{A} For the subset of subjects (n=17) with at least one seizure in the typical and tapered condition and their 304 seizures, we plotted seizure duration after removing systematic differences between subjects (using a random offset linear mixed-effects model -- LMM) against normalised ASM levels. Seizures occurring in the typical and tapered conditions are shown in orange and pink, respectively. Seizures containing taper-emergent states are highlighted in blue. Best fit of fixed effects from a LMM is shown as a solid line, with a slope $\beta = -0.058$, $p < 0.001$.
    \textbf{B} For the subset of subjects (n=7) that had at least one seizure containing taper-emergent states, we additionally show their seizure duration after removing subject-level difference for three categories as a box plot. 
    The three categories are from left to right: Seizure during tapered condition, containing taper-emergent states (blue); Seizure during tapered condition, without taper-emergent states (pink); Seizure during typical ASM condition (orange).
    \textbf{C} For the subset of n=7 subjects and their 30 seizures that contained taper-emergent states, we scattered percentage of seizure duration in taper-emergent state against normalise ASM levels. Each data point is a seizure, each colour is a subject.
    }
    \label{fig:result_dur_by_state}
\end{figure}

Given these taper-emergent states, we investigated whether their presence impacted the duration of seizures. Using the 304 seizures across 17 subjects from the previous section, which ensured that each subject had at least one seizure arising from the typical and tapered condition, we first sought to confirm the correlation between seizure duration and normalised ASM levels as before (Figure~\ref{fig:result_dur_by_state}A). Using linear mixed-effects regression, we obtained $\beta = -0.058$, $p < 0.001$ as the coefficient for ASM levels in predicting seizure duration. This translates to an approximate 12.5\% increase in seizure duration for every unit reduction in normalised ASM levels, which is within our previous estimation bounds. Figure~\ref{fig:result_dur_by_state}A additionally highlights those seizures containing taper-emergent states (as blue data points), and visually we can already see that these seizures tended to be prolonged.

We then isolate the seven subjects that had seizures containing taper-emergent states, with all their 80 seizures in total (30 seizures with taper-emergent states and 50 seizures without). We compared log-durations between seizures with and without taper-emergent states (Figure~\ref{fig:result_dur_by_state}B), confirming that seizures containing taper-emergent states tended to be longer than those without, and substantially longer (Wilcoxon rank-sum $r=0.485$, $p < 0.001$) than seizures occurring during the typical ASM condition. To better account for subject-level effects statistically, we also tested this relationship using a mixed-effect model ($logDuration \sim seizureCategory + (1|subject)$, where $seizureCategory$ corresponds to the three categories in Figure~\ref{fig:result_dur_by_state}B) and found that both groups of seizures without taper-emergent states ($\beta_{Tap}=-0.178, p=0.022$ for tapered condition, and $\beta_{Typ}=-0.288, p < 0.001$ for typical condition) are substantially shorter than seizures with taper-emergent states. Again, note that $\beta_{Tap}=-0.178$ indicates a 66\% decrease in seizure duration of seizures in the tapered condition without the taper-emergent states relative to seizures with.

Finally, we determined the proportion of the seizure duration that was occupied by taper-emergent states in those 30 seizures and seven subjects that contained them. We found a median proportion of 38.5\% across subjects and seizures, and these taper emergent states tended to occur later on in the seizure (Suppl.~\ref{SUP_TpaSS}). However, the proportion of seizure duration taken up by the taper-emergent states was not correlated with ASM level across all subjects, but instead followed an all-or-nothing pattern in each subject (Figure~\ref{fig:result_dur_by_state}C).

\subsection{Tapering also prolongs seizures without taper-emergent states}

Finally, we tested if seizures without taper-emergent states were also prolonged with medication tapering. That is, we tested whether seizures consisting purely of dose-independent states were also prolonged with medication tapering. We included all seizures from the previous analysis (all subjects that had at least one seizure in each tapered and typical ASM condition) and removed any seizures containing taper-emergent states (resulting in 274 seizures across 17 subjects).

\begin{figure}[H]
    \centering
    \includegraphics[scale=1]{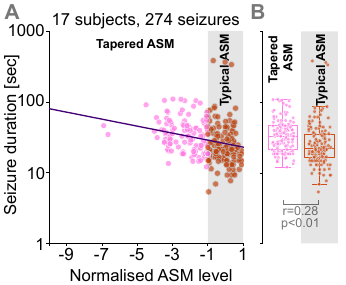}
    \caption{
    \textbf{Seizures without taper-emergent states are also prolonged with tapering.}
    \textbf{A} For the subset of subjects (n=17) with at least one seizure in the typical and tapered condition and their 274 seizures that did not contain taper-emergent states, we plotted seizure duration after removing systematic differences between subjects (using a random offset linear mixed-effects model -- LMM) against normalised ASM levels. Seizures occurring in the typical and tapered conditions are shown in orange and pink, respectively. Best fit of fixed effects from a LMM is shown as a solid line, with a slope $\beta = -0.049$, $p < 0.001$.
    \textbf{B} For the same subjects and seizures, we additionally show their seizure duration after removing subject-level difference as a box plot for tapered and typical ASM conditions.
    }
    \label{fig:result_asm_cmn}
\end{figure}

Figure~\ref{fig:result_asm_cmn}A demonstrates, as expected, that lower ASM-levels were associated with longer seizures ($\beta = -0.049$, $p < 0.001$) in this subset. When comparing tapered \textit{vs.} typical ASM condition directly (Figure~\ref{fig:result_asm_cmn}B), we found that seizures occurring during ASM tapering were substantially longer than those occurring in the typical condition (Wilcoxon rank-sum: $r=0.275$, $p < 0.001$). Both results indicate that ASM tapering also prolongs seizures without taper-emergent states.

It is important to note that this result does not mean that all seizures without taper-emergent states show the same seizure patterns and sequences, and are simply elongated, or ``stretched out'' evenly with tapering. Rather, we found that tapering makes some states more common in seizures, and that only one or few states are elongated in each subject, resulting in an overall longer duration (Supplementary~\ref{SUP_TypSS}).

\newpage

\section{Discussion}
\label{sec:Discussion}

Our study provides new insights into how anti-seizure medications (ASMs) modulate seizure activity, revealing two distinct mechanisms by which they influence seizure duration. First, we found that ASM withdrawal can lead to the emergence of new seizure activity patterns, which were associated with significantly longer seizures in approximately 40\% of patients. Second, even in the absence of these newly emergent states, lower ASM levels still resulted in prolonged seizure durations, with increases ranging from 12\% to 224\%, depending on the extent of tapering. These findings suggest that ASMs not only compress seizure duration but also selectively suppress specific seizure patterns in an all-or-nothing manner. This dual role of ASMs in seizure control highlights their complex impact on brain activity and underscores the need for personalized dosing strategies to optimize epilepsy management.

Our findings have important implications for drug resistance in epilepsy. Previous research has suggested that ASMs shorten seizures, leading to the idea that they work by simply `compressing' seizure activity into shorter durations. Our findings challenge this notion: We found that specific brain regions and seizure activity patterns emerge at certain ASM thresholds (Fig.~\ref{fig:result_dur_by_state}), aligning with clinical observations that seizure suppression occurs abruptly rather than gradually with ASM titration. We propose that this all-or-nothing suppression reflects a \textit{dynamic bifurcation}, where a gradual change in ASM levels triggers a sudden shift in seizure pattern. In contrast, the gradual shortening of seizures represents a continuous, parameter-driven change. While the latter is expected in most (biological) systems, the former is rare — and linked to critical transition theory. This theory has been proposed for the onset of seizures \citep{da2003epilepsies,jirsa2014nature,lepeu2024critical}, but our key contribution is demonstrating that these critical transitions also independently apply to other seizure activity patterns with ASM tapering. Future studies should systematically investigate how different ASM types induce these mechanisms, considering variability across brain regions and epilepsy aetiologies. Ultimately, this could lead to a predictive model, where we observe the different seizure patterns arising in a given patient, and can predict the optimal ASM type and dosage needed to fully suppress all of them.

We demonstrate how ASM taper-emergent seizure dynamics may involve ``new'' brain regions that were spared under typical ASM conditions. This highlights how a mapping of the interaction between these regions and ASM levels could provide important clues into seizure suppression, and drug-resistance. Further, a recent functional magnetic resonance imaging (fMRI) study of participants who had recently experienced their first seizure,~\cite{pedersen2024brain} revealed changes in widespread brain network function following ASM administration. These findings suggest that ASMs may impart their effects by engaging a spatial network of brain regions. In the context of our results, it highlights the question if specific brain networks experience one ASM mechanism more so than another. This question also needs to be answered in the spatial context of the seizure origin. In our study, the sample size was insufficient to test this, but future work should be able to shed more light on this matter.

To provide some further interpretation of these seizure states, we highlight a particular area for future research. These seizure states correspond to specific activity patterns in specific brain regions. They will most likely correlate with seizure semiology states and networks \citep{mcgonigal2021seizure}, and could be formally tested in future work. Importantly, in some patients we may not be able to suppress all seizure states with ASMs, but may need to focus on some states that are more challenging, e.g., as they represent particular symptoms (loss of awareness or similar). Clinically, we know that even in drug-resistant patients, ASMs help suppress particularly severe seizures, most likely because they rescue critical function and -- in the language of our work -- effectively prevent certain seizure states from being accessed. Future work should also test if particular seizure states influence the post-ictal state and recovery. If so, these states could become targets for ASM treatment. In parallel, we should also investigate if ASMs have a direct effect on the post-ictal recovery independent of seizure states.

Furthermore, we speculate if a similar dual mechanism is at play for interictal brain activity states. Patients on ASMs often experience cognitive impairments as a result of the medication \citep{dusanter2023cognitive}, and it would be interesting to assess if this may be due to certain brain states suddenly becoming inaccessible. Similarly, ASMs are thought to also impact sleep and sleep architecture \citep{liguori2021effects}. We speculate that the dual mechanism shown here has relevance for understanding the relationship of certain ASMs and sleep, independent of seizure occurrence. Finally, ASMs are also used for other conditions such as mood disorders and migraines. Future work should investigate the relevance of either mechanism in those conditions.

\textit{Limitations and future work}: Our study was performed on a retrospective, observational, small sample. This means that a lot of questions around causality and covariates could not be answered definitively. For example, we found no evidence of any specific types of epilepsy or ASM influencing our results. This does not mean that there is no effect, but simply that we require larger samples to investigate the size and direction of these covariates and interactions. A few technical caveats should also be noted. We used electrographic seizure duration here, as our whole analysis rested on the electrographic data. However, clinical seizure duration may differ and future work should explore implications thereof. Our patient cohort are all invasively implanted with iEEG, meaning that on the first day of recording anaesthesia will add on top of ASM effects. We did not have sufficient sample size to investigate this in more detail, but it is possible for future studies to disentangle any effects. We were also not able to directly investigate the effect of sleep, as this data was not consistently recorded in our retrospective data. Our circadian analysis suggests that there is a subtle effect across the cohort in terms of seizure duration, which we account for in our main results (see Suppl.~\ref{SUP_MELM}). However, any future work with larger cohorts should account for sleep and chronobiological effects, as we know seizures and seizure states are influenced by these \citep{schroeder2020seizure,panagiotopoulouFluctuationsEEGBand2022,karoly2021cycles,Schroeder2023}.

\newpage

\section{Acknowledgments}
We thank members of the Computational Neurology, Neuroscience \& Psychiatry Lab (www.cnnp-lab.com) for discussions on the analysis and manuscript.

We further thank Charlotte McLaughlin and all the NHNN Telemetry Unit staff, and are grateful to all the people with epilepsy whose intracranial EEG data enabled this research.

\section{Funding}
P.N.T. and Y.W. are both supported by UKRI Future Leaders Fellowships (MR/T04294X/1, MR/V026569/1). JSD, JdT are supported by the NIHR UCLH/UCL Biomedical Research Centre. S.J.G is supported by the Engineering and Physical Sciences Research Council (EP/L015358/1) and ADLINK. N.E is supported by a Epilepsy Research Institute (ERI) Studentship.

\section{Competing interests}
The authors report no competing interests.

\newpage

\bibliography{ref}
\newpage


\newpage
\renewcommand{\thefigure}{S\arabic{figure}}
\setcounter{figure}{0}
\setcounter{equation}{0}
\counterwithin{figure}{section}
\counterwithin{equation}{section}
\counterwithin{table}{section}
\renewcommand\thesection{S\arabic{section}}
\setcounter{section}{0}

\section*{Supplementary}

\section{Subjects details}
Table~\ref{table:Demograph} provides details regarding demographics, iEEG recordings, seizure types and tapered ASMs. We provide these details for the complete dataset, the reduced dataset, and the seven individuals with taper-emergent states. Remember in the reduced dataset, we selected subjects who had at least three seizures overall, with at least one in each of the tapered and typical ASM conditions. In the seven individuals with taper-emergent states, we did not find any single medication specific to these subjects.

\begin{table}[H]
\centering
\begin{tabular}{ l | l  l l} 
                                 & Complete       & Reduced dataset & ``Taper-emergent''\\
\hline
No. subjects                     & 28             & 17              & 7\\
Age (mean$\pm$SD)                & 33.1$\pm$8.2   & 33.0$\pm$7.3    & 33.9$\pm$8.2\\ 
Sex (F\%)                        & 46.4\%         & 29.4\%          & 0.0\% \\
Epilepsy (TLE\%)                 & 57.1\%         & 64.7\%          & 57.1\%\\
Side (Left\%)                    & 39.3\%         & 35.3\%          & 42.9\%\\
No. iEEG contacts (mean$\pm$SD) & 86.9$\pm$23.2  & 82.1$\pm$20.2   & 91.4$\pm$22.7\\
No. ROIs (mean$\pm$SD)           & 19.6$\pm$5.5   & 18.5$\pm$5.0    & 18.0$\pm$2.9\\
Days of recording (mean$\pm$SD)  & 12.2$\pm$2.7   & 12.3$\pm$3.2    & 13.1$\pm$2.3\\
No. seizures                     & 457            & 308             & 80\\
\MyIndent Sub-clinical           & 45.7\%         & 43.5\%          & 36.2\%\\
\MyIndent Focal                  & 52.1\%         & 54.3\%          & 60.0\%\\
\MyIndent SG                     & 0.7\%          & 0.6\%           & 2.5\%\\
\MyIndent Other                  & 1.5\%          & 1.6\%           & 1.2\%\\
No. regular ASM (mean$\pm$SD)    & 3.2$\pm$1.0    & 3.2$\pm$1.0     & 3.1$\pm$0.3\\
No. tapered ASM (mean$\pm$SD)    & 2.3$\pm$0.7    & 2.3$\pm$0.7     & 2.7$\pm$0.5\\

\end{tabular}
\caption{Summary of subject data used in the analysis. SG: secondarily generalized Seizures. SD: Standard Deviation, F: Female, TLE: Temporal Lobe Epilepsy, ROI: Region Of Interest, ASM: Anti-Seizure Medication, iEEG: Intra-Cranial Electroencephalography}
\label{table:Demograph}
\end{table}

\section{Algorithmics details on seizure states identification}\label{SUP_2}
\subsection{Pre-ictal period noise removal}\
As stated on \textit{Methods, Identification of seizure states}, we used the pre-ictal period (commencing 120 seconds prior to seizure onset) as a baseline to z-score band-power during the ictal period. We excluded potential preictal noise by removing any preictal time-window with a Median Absolute Deviation (MAD) score exceeding a threshold of five. For each combination of region and frequency band, we z-scored ictal values against the preictal distribution. Any negative ictal z-scores were excluded (i.e., replaced with 0) as we were only interested in increases in activity compared to the preictal segment. Note that negative ictal z-scores most likely represent elctrodecrement events, which should be treated differently to seizure activity, and we reserve their investigation for a separate paper.

\subsection{Reducing observation for Non-negative Matrix Factorisation (NMF)}\
We aimed to avoid the NMF algorithm over-prioritising states that a seizure stays in for a extended period of time. This was achieved by clustering time points in the data matrix $X$ if they had a correlation coefficient of more than 0.8. We then kept only one time point in each cluster to produce a reduced matrix $rX$ with dimensions $(n_{regions}*n_{bands}) \times
n_{T_{reduced}}$, where $n_{T_{reduced}}$ is the number of clusters below  0.8 in correlation coefficient.

We applied NMF as described in the main text to this reduced data matrix to obtain: $rX \approx W \times rH$ and reconstructed the `actual' $H$ matrix by $H = W \textbackslash X$, where $X$ is the original data matrix. We essentially used the $W$ matrix from the reduced data matrix to reconstruct what $H$ must look like given the original data matrix $X$. Note this $H$ matrix is not guaranteed to be non-negative now, but will contain small negative numbers, which we will understand as ``noise'' and leverage later.

\subsection{NMF optimisation}\
When running the NMF algorithm, we selected a value for the number of components ($n_{c}$) by testing a range of values from 2 to $\sqrt{n_{T_{reduced}}}$ and finding the value for $n_{c}$ that produced the most stable components. This was tested following the steps shown in \citesupp{brunetMetagenesMolecularPattern2004, kimSparseNonnegativeMatrix2007}. For each value of $n_{c}$ the NMF algorithm was ran 30 times and an average connectivity matrix C was produced. We then calculated the dispersion coefficient $\rho$ using:
\begin{equation}
 \rho = \frac{1}{n^{2}}\sum_{i=1}^{n}\sum_{j=1}^{n}*\left((C_{ij}-\frac{1}{2}\right)^2
\label{SuppRHO}
\end{equation}
We then selected the value for $n_{c}$ that maximised $\rho$.

\subsection{Definition of null state}
As explained in methods section ~\ref{METH_2.4}, each time window is assigned to the most strongly expressed component to create the states analysed in this paper. However if no component is expressed strongly, a time point will be assigned to a ``null'' state. To define the threshold for low expression, we found an estimation of the "noise" in $H$. We leveraged the fact that the $H$ matrix contains small negative numbers, which we assume to be ``noise'' from the NMF reduction. We mirror these negative numbers to give us a symmetric distribution around zero, derive the standard deviation of said distribution, and set the threshold of the null state at 3 times the standard deviation. We are effectively saying that -- assuming the small negative and positive numbers in $H$ are noise -- we only deem a time window to be a seizure state if it exceeds this noise threshold. Note that we are most likely ignoring small seizure activity limited to few channels and frequency bands (e.g. at the start of seizures). Such onset patterns would most likely fall below the threshold and be classified as a null state. Future work wishing to investigate onset patterns specifically will have to modify our algorithm. In this work, we were not interested in onset patterns, but were only investigating seizure propagation patterns.

\section{Hierarchical Models to determine relationship between seizure duration and ASM levels}\label{SUP_MELM}

To determine the relationship between normalised ASM levels and seizure duration, a linear mixed-effects model (LMM) was used:\\ 
\begin{equation}
    \begin{split}
    log{\text -} duration \sim &  \beta_0 \\
    & + \beta_1ASM{\text -}level \\
    & + \beta_2sin(TOD) + \beta_3cos(TOD) \\
    & + \beta_4sin(12hTOD) + \beta_5cos(12hTOD) \\
    &+ (1|subject)
    \end{split}
\label{SuppMELM}
\end{equation}

Duration on log10 scale  ($log{\text -} duration$) was predicted using a fixed-effect on normalized ASM plasma concentration levels ($ASM\_lvl$) and a random intercept for each subject ($subject$). To account for circadian and 12 hour fluctuations fixed-effect on time of day (TOD) was included as a circular variable. Sine ($sin(TOD)$) and cosine ($cos(TOD)$) of time of day represent circadian rhythms, while $sin(12hTOD)$ and $cos(12hTOD)$ represent the ultradian rhythm 12~hour cycle. One reason for including the 12~hour cycle is to factor out any effects of ASM intake which are 12-hourly (morning and evening intake).

This model was applied in three scenarios: 1) the complete dataset of 28~individuals and 457~seizures, 2) reduced dataset for seizure states analysis of 17~individuals and 308~seizures, 3) same 17 individual, but without seizures containing taper-emergent states resulting in 274~seizures. Coefficients and p-values for these three cases are reported on Table~\ref{table:melm} including fixed-effects of ASM plasma levels.

\begin{table}[h!]
\tiny
\centering
\begin{tabular}{|l  c | c c c c c|} 
 \hline
 
Dataset [subj., sz.]    &             & ASM-lvl        & sin(TOD)& cos(TOD)& sin(12hTOD)& cos(12hTOD)\\ [0.5ex] 
\hline
Full Dataset [28, 457]     & $\beta$ & -0.032         & -0.057  & -0.035  & -0.026     & -0.015\\
                           & p       & \textless0.001 & 0.008   & 0.133   & 0.252      & 0.474 \\
\hline
States Study [17, 304]     & $\beta$ & -0.058         & -0.049  & -0.012  & -0.011     & -0.041\\
                           & p       & \textless0.001 & 0.044   & 0.644   & 0.664      & 0.086 \\
\hline
Typical Seizures [17, 274] & $\beta$ & -0.049         & -0.058  & -0.029  & -0.028     & -0.049\\
                           & p       & \textless0.001 & 0.016   & 0.296   & 0.285      & 0.044 \\
\hline
 
\end{tabular}
\caption{Summary of the hierarchical models in eqn.~\ref{SuppMELM} fitted to generate the main results of this study showing the fixed-effect coefficients ($\beta$) and p-values (p), including ASM levels and the circular variables representing the circadian (time of day - TOD) and 12h (12hTOD) effects.}
\label{table:melm}
\end{table}

\subsection{Interaction of ASM levels and time of day}
\begin{figure}[H]
    \centering
    \includegraphics[scale=0.8]{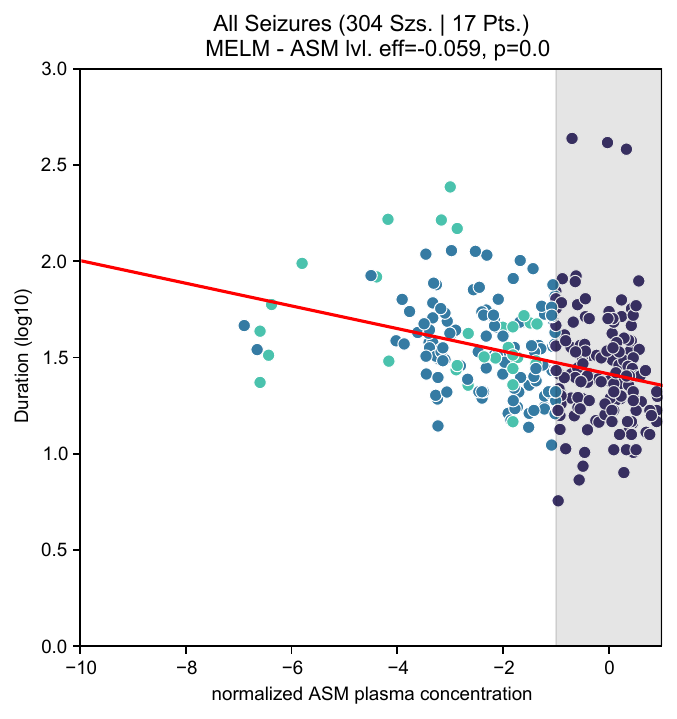}
    \caption{\textbf{Exclusion of time of day does not distort the effect of ASM in predicting seizure duration.} We reproduced Figure~\ref{fig:result_dur_by_state}A without the time of day circular variables.\\
    For the subset of subjects (n=17) with at least one seizure in the typical and tapered condition and their 304~seizures, we plotted seizure duration after removing systematic differences between subjects (using a random offset linear mixed-effects model -- LMM) against normalised ASM levels. Compared to the LMM presented in methods, this new model does not include fixed-effect related to 24h and 12h rhythms. Seizures occurring in the typical and tapered conditions are shown in navy and blue, respectively. Seizure containing taper-emergent states are highlighted in cyan. Best fit of fixed effects from a LMM is shown as a solid red line, with a slope $\beta = -0.059$, $p < 0.001$.}
    \label{fig:sup_asm_sss_ntChrn}
\end{figure}

We next sought to test if excluding the time of day (both 24h and 12h) effects impacts our findings of ASM levels. To this end, we fitted a new model only featuring the ASM levels and the random-effect of subject: [$log{\text -} duration \sim \beta_0 + \beta_1ASM{\text -}level + (1|subject)$]. Figure~\ref{fig:sup_asm_sss_ntChrn} mimics Figure~\ref{fig:result_dur_by_state}A with the modified model, showing no substantial change on the new model. $\beta_1$ shows a difference between models around 0.001 (representing a difference of 0.2\% on increase rate) with similar p-values ($< 0.001$).

\section{Further characterizing taper-emergent states}\label{SUP_TpaSS}
We performed additional analyses characterizing taper-emergent states. As stated in our methods, seizures with taper-emergent states were longer compared to seizures without them ($p < 0.001$). 

\begin{figure}[H]
    \centering
    \includegraphics[scale=0.8]{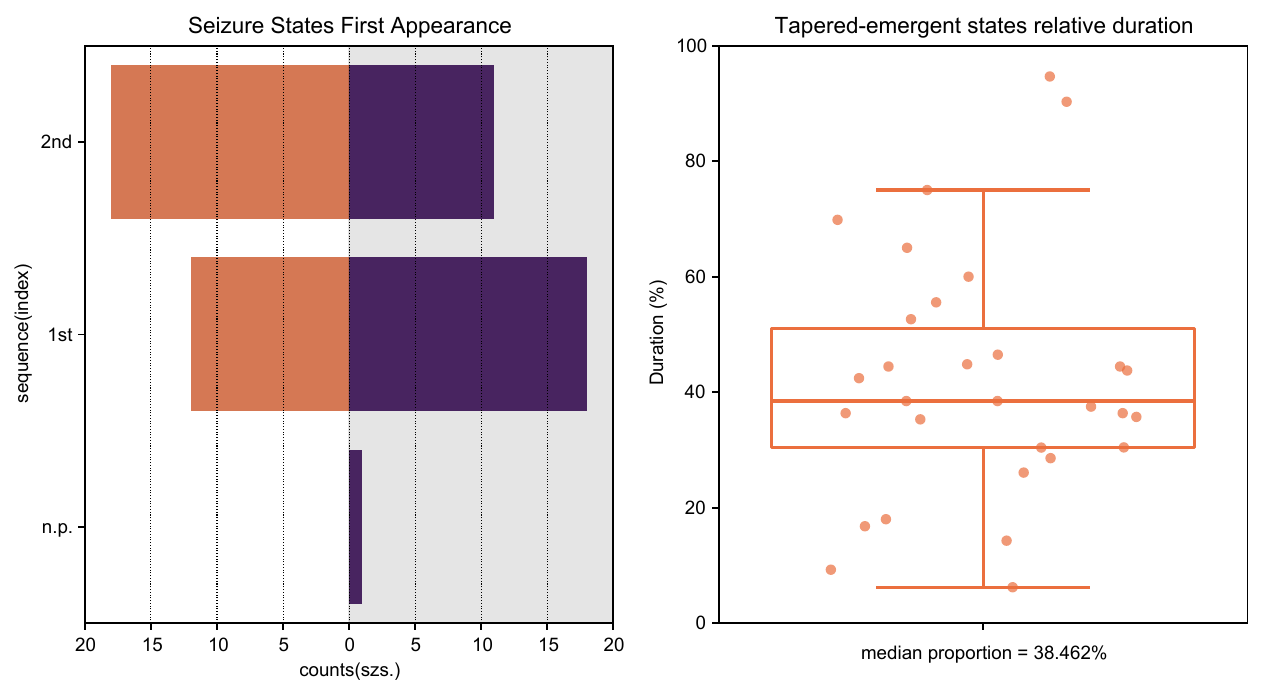}
    \caption{\textbf{Taper-emergent states have a substantial contribution to seizure duration predominately during later stages of the seizure}. \textbf{Right}: Study of states sequence comparing taper-emergent states (orange) and dose-independent states (purple) on seizure containing tapered-emergent states (30 seizures, 7 individuals). The bar-plot represent the number of seizures where the state type is present the first (1st), the second (2nd) or it is not present (n.p.).}
    \label{fig:SUP_TpaSS_perc_seq}
\end{figure}

We also studied the sequence of states within taper-emergent seizures (Figure~\ref{fig:SUP_TpaSS_perc_seq}~left) comparing the first appearance of each state type. In 60\% of seizures with taper-emergent states, these appeared as the second or later in the sequence of states.

\begin{figure}[H]
    \centering
    \includegraphics[scale=0.45]{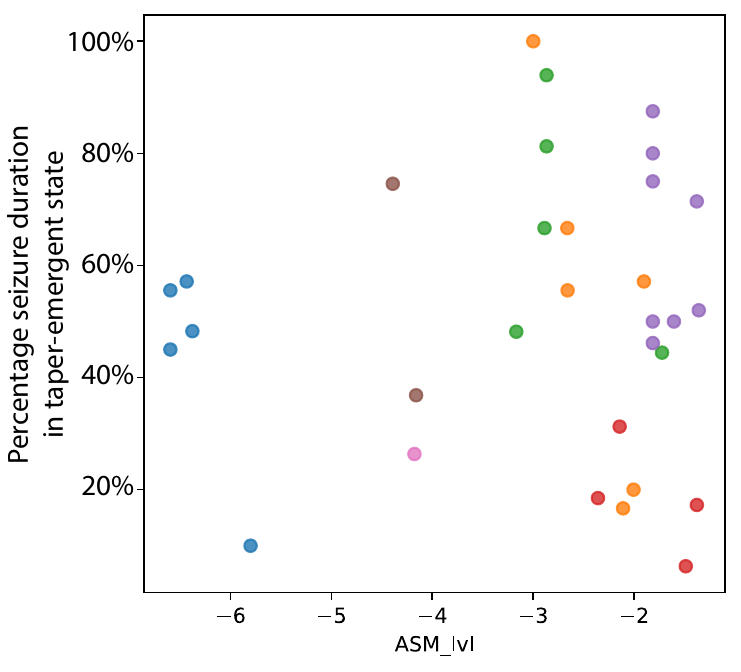}
    \caption{\textbf{Contribution of taper-emergent states to overall seizure duration plotted against ASM levels.} Each data-point corresponds to a seizure (30~seizures) colour-coded by individuals (n=7).}
    \label{fig:sup_perASM}
\end{figure}

Figure~\ref{fig:SUP_TpaSS_perc_seq}~right further shows that taper-emergent states substantially contribute to overall seizure duration, with a median across these seizures (N=30) of 38.5\%. Moreover, this contribution was not correlated with ASM level across all subjects (Figure~\ref{fig:sup_perASM}), but followed more a all-or-nothing pattern: they emerge suddenly at a subject-specific normalised ASM level and can contribute to between 20-80\% of the overall seizure duration.

Finally, we compared signal amplitude and spatial spread between taper-emergent and dose-independent states. As a result of the NMF, the data matrix for each individual ($X = (nROI*nFrq) \times nEpoch$) was decomposed into matrices W ($nROI*nFrq \times nStt$) and H ($nStt \times nEpoch$). The W matrix can be interpreted as levels of band power involvement (abnormal relative to baseline, i.e. involved in seizure) of each state across ROIs and frequency bands. To identify which ROIs were recruited by each state, the following steps were followed: 
\begin{enumerate}
    \item W across frequency bands for each ROI is squared and summed across frequency bands for each ROI.
    \item The resulting number is dived by the number of ROIs.
    \item Recruited ROIs are identified if the final number exceeds~1.
\end{enumerate}
As each state in $W$ is of euclidean norm 1, the above steps measure above average involvement of each channel.

\begin{figure}[H]
    \centering
    \includegraphics[scale=0.45]{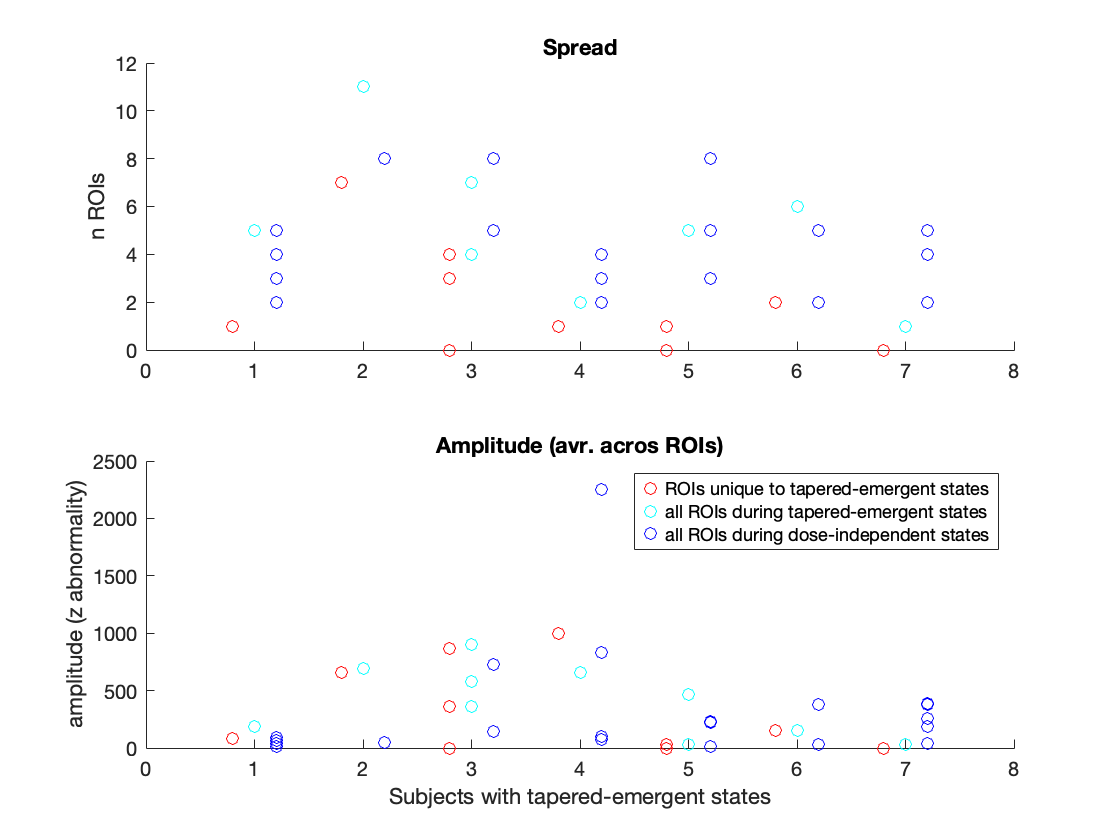}
    \caption{\textbf{Taper-emergent states recruit additional ROIs, and show higher amplitude than dose-independent states}. For both panels each data-point represents a states. X-axis numbers are separate subjects. Dose-independent states are shown in navy blue. Taper-emergent states are presented in cyan. Red data points show regions that were exclusive to taper-emergent states. \textbf{Top}: Seizure state spread measured as the number of ROIs with involvement in said state. \textbf{Bottom}: Average amplitude (z-score relative to preictal) of the ROIs for each state. }
    \label{fig:sup_states_amplt}
\end{figure}

From this, we identified ROIs that were recruited by each state type (taper-emergent and dose-independent), as well as which ROIs were exclusively recruited by taper-emergent states. Taper-emergent and dose-independent states do not show any mayor differences on the number of recruited ROIs (Figure~\ref{fig:sup_states_amplt}A). However, we found at least one additional ROI is exclusively recruited during taper-emergent states in most subjects (6/7). With the recruited ROIs identified, the amplitude during taper-emergent state can be compared.

Amplitude for each ROI was obtained as the sum of the squares of the band power for each frequency bands. Note that amplitude is to be interpreted as the abnormality z-score of the ictal log band power compared to pre-ictal. Then, for each state we averaged across recruited ROIs and active time-window (Figure~\ref{fig:sup_states_amplt}B). Additionally, the average amplitude for ROIs exclusively recruited by taper-emergent states is estimated. Taper-emergent states show a higher amplitude compared to dose-independent states in some subjects (4/7), despite ROIs exclusively recruited by taper-emergent state showing lower amplitude than the average.

\section{Further characterizing dose-independent states}\label{SUP_TypSS}
We further characterize seizures only composed of dose-independent states (274~seizures, 17~individuals). As a reminder, dose-independent states appear in both conditions of typical and tapered ASM. In other words, we have removed any seizures that contain taper-emergent states for this analyis. 

\begin{figure}[H]
    \centering
    \includegraphics[scale=0.9]{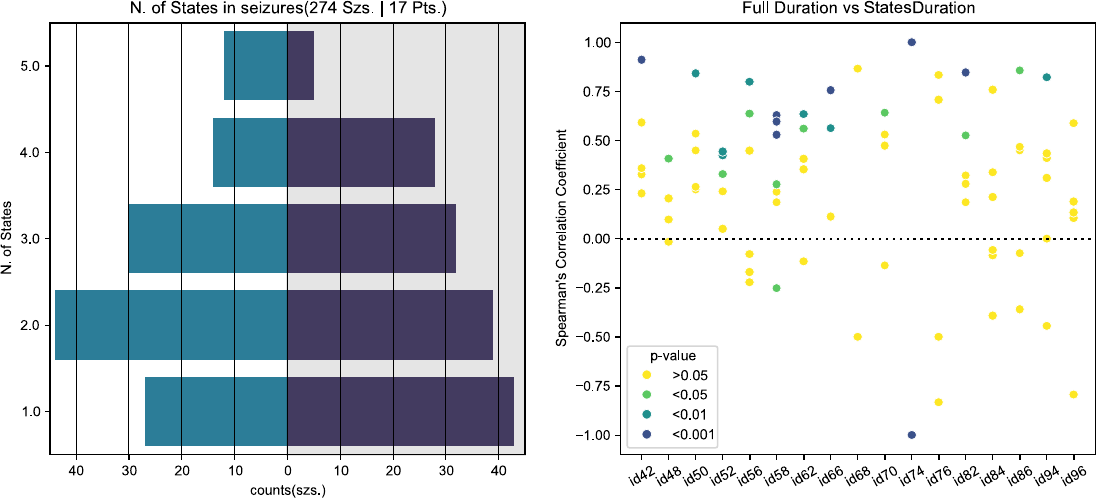}
    \caption{\textbf{The increase in duration of seizures without taper-emergent states is not homogeneously distributed amongst the states}. This study is applied to all seizures without taper-emergent states, i.e. solely consisting of dose-independent states (274~seizures, 17~individuals). \textbf{Left}: Distribution of the number of states in each seizure, summarised across all seizure for the typical (right, navy colour) and tapered (left, blue colour) condition. \textbf{Right}: Correlation between individual state duration and the overall seizure duration. Each point represent a state, sorted by subjects in columns. The y-axis indicates the Spearman's Correlation Coefficient between state duration and seizure duration, and the colour of each data point is based on p-values.}
    \label{fig:SUP_TypSS_corr_Nstt}
\end{figure}

Analysing the number of states in these seizures (Figure~\ref{fig:SUP_TypSS_corr_Nstt}A), there is no difference between seizures occurring during typical and tapered conditions.

Given that multiple dose-independent states were present in each seizure (average: 2.5, standard deviation: 1.2), the contribution of these states to the total seizure duration was studied. We assessed if all dose-independent states in a seizure were contributing equally to total seizure duration. For every state we applied Spearman Correlation between its duration and the total seizure duration (Fig.~\ref{fig:SUP_TypSS_corr_Nstt}B) across all seizures in each subject. We identified at least one state with a correlation coefficient above 0.7 and $p < 0.05$ in most subjects, but by no means were all state durations highly correlated with overall seizure duration. This indicates that the prolonged seizure duration with ASM tapering is by no means ``stretching'' all states equally (we would otherwise see perfect 1:1 correlations in all states). Rather, in most subjects, one or few states are stretched in line with overall seizure duration. In other subjects, there is no consistent pattern.

\bibliographystylesupp{plainnat}
\bibliographysupp{supplementary}

\end{document}